\documentclass[12pt,a4paper]{article}
\usepackage{amssymb,a4wide,xypic}

\newtheorem{thm}{Theorem}[section]

\newtheorem{lemma}[thm]{Lemma}
\newtheorem{prop}[thm]{Proposition}
\newtheorem{defin}[thm]{Definition}
\newtheorem{assumption}[thm]{Assumption}

\newtheorem{remark}[thm]{Remark}
\newtheorem{example}[thm]{Example}
\newenvironment{proof}{\par\smallskip\rm\emph{Proof.}}{\qed}
\newcommand{\qed}{\hfill$\square$\par\smallskip}

\newcommand{\g}{{\mathfrak g}}
\newcommand{\Mg}{{\mathfrak M}}
\newcommand{\Cc}{{\mathcal C}}
\newcommand{\E}{{\mathcal E}}
\newcommand{\A}{{\mathcal A}}
\newcommand{\Ag}{{\mathfrak A}}
\newcommand{\Z}{{\mathbb Z}}
\newcommand{\R}{{\mathbb R}}
\newcommand{\C}{{\mathbb C}}
\newcommand{\Lie}{{\mathcal L}}
\newcommand{\Mc}{{\mathcal M}}
\newcommand{\Nc}{{\mathcal N}}
\newcommand{\beq}{\begin{equation}}
\newcommand{\eeq}{\end{equation}}
\newcommand{\beqa}{\begin{eqnarray}}

\newcommand{\vf}[2]{{\frac{\partial #1}{\partial #2}}}
\newcommand{\ber}{{\mathcal Ber}}

\begin{document}
\title{Superlocalization  formulas and supersymmetric Yang-Mills theories}
\author{U. Bruzzo \\[-3pt]   \normalsize
Scuola Internazionale Superiore di Studi Avanzati, \\[-6pt]   \normalsize Via Beirut 4, 34013 Trieste, Italy
\\[-6pt]  \normalsize and I.N.F.N.,    Sezione di Trieste
\and
F. Fucito \\[-3pt]    \normalsize
Dipartimento di Fisica, Universit\`a di Roma ``Tor Vergata'', and
I.N.F.N., \\[-6pt]   \normalsize Sezione di Roma II,
Via della Ricerca Scientifica,  00133 Roma, Italy}
\maketitle\thispagestyle{empty}
\begin{abstract}
By using  supermanifold techniques we prove a generalization
of the localization formula in equivariant cohomology which is
suitable for studying  supersymmetric Yang-Mills theories in terms of ADHM data.
With these techniques one can compute the reduced partition
functions of topological super Yang-Mills theory with 4, 8 or 16
supercharges. More generally, the superlocalization formula 
can be applied  to any topological field theory in any number of dimensions.
\end{abstract}
\newpage\setcounter{page}{1}
\section{Introduction}
The study of nonperturbative effects in nonabelian
supersymmetric gauge theories (SYM) has been the focus of much
research activity in theoretical physics. In recent years, after
the work of Seiberg and Witten
\cite{Seiberg:1994rs,Seiberg:1994aj}, many new results have been
obtained and the techniques discussed in
\cite{Seiberg:1994rs,Seiberg:1994aj} have also been used to shed
light on analogous effects in string theory (see 
\cite{Lerche:1997sm} for a review). Even if these techniques are
very powerful, yet the derivation of the above mentioned results
relies on a certain number of assumptions. This has triggered the
interest of some authors to check these results by independent 
methods. A direct evaluation of nonperturbative effects is in
fact possible by computing the partition function or the relevant
correlators of the theory of interest. It has been a pleasant
surprise of the last year that such a computation can indeed be
carried out in some cases for arbitrary values of the instanton winding number. 

This is the end point of
a long journey that we will very briefly summarize: following the
pioneering works \cite{Amati:1988ft,Shifman:1994ee} in which
condensates for SYM theories were computed, in
\cite{Dorey:1996hu,Dorey:1996bf,Fucito:1997ua} a first check for
the prepotential of the SYM with eight fermionic charges was
performed for instantons of winding number two. This computations
were performed at the semiclassical level and agreed with the
prepotential of \cite{Seiberg:1994rs,Seiberg:1994aj}. This opened
the way to a reformulation in the framework
of topological theories \cite{Bellisai:2000tn,Bellisai:2000bc}. 
The full exploitation of the  power of this
observation was hindered  by the presence of constrained quantities in the
functional integral (see \cite{Dorey:2002ik} for a detailed
account on this point). After this point was dealt with, it was
possible to show that the measure of the functional integral could
be written as a total derivative \cite{BFTT}. The derivative
operator is found to be the BRST operator of the SYM. Even if the
computation was then reduced to the evaluation of a boundary term
there were still some issues to be settled concerning the
compactness of the domain of integration. This was done for the
first time in the case of instantons of winding number two
\cite{Hollowood:2002ds,Hollowood:2002zv}, by localization on
certain manifolds of given dimensions. The computation for
arbitrary winding number in the case of eight supersymmetries with
matter in the fundamental and adjoint representation was presented
in \cite{Nekrasov:2002qd}. In turn this work stemmed from a line
of research focused on computing topological invariants with the
aid of topological field theories
\cite{Losev:1998tp,Moore:1997dj,Moore:1998et}. 

An interesting
alternative approach to the computation in \cite{Nekrasov:2002qd}
appeared in \cite{Flume:2002az}, where use was made of the
localization formula in equivariant cohomology (for
this formula see e.g.  \cite{BGV}). In order to use that formula, the action of the SYM under study
must be reinterpreted as a form over the instanton moduli space. This is quite natural for SYM theories with eight
charges \cite{Witten:1988ze}, where the fermionic moduli
can be interpreted as differential forms on the instanton moduli
space. In the case of instantons of
arbitrary winding number, this approach can be conveniently
implemented by using the ADHM contruction \cite{ADHM}, as it was shown in
\cite{Bellisai:2000tn,Bellisai:2000bc}.

In this paper we want to show how by using supermanifold
techniques one can write a (mild) generalization of the
localization formula which is suitable for implementing
this approach to the computation of the partition function
to SYM theories with any number of charges (supermanifold techniques
were already used in \cite{BFTT} to retrieve the measure of
the partition function of SYM). The structure of the paper is as
follow. In Section \ref{Sec2} we briefly recall the basics
of equivariant cohomology and the localization formula.
In  Section \ref{Sec3} we spell out what are the bundles on the instanton
moduli space that are relevant to different numbers of supersymmetry
charges. In  Section \ref{Sec4} we review the basics of supermanifold theory,
paying special attention to the supermanifolds defined in terms
of the cotangent bundle to a differentiable manifolds (the ``tautological
supermanifolds''). In  Section \ref{Sec5} we describe our
generalization of the superlocalization formula; this stems
from the identification of the BRST operator with a suitable
equivariant cohomology operator, which in the case of
tautological supermanifolds (i.e., in the $\Nc=2$ case) is
the usual equivariant exterior differential. 
Finally in Section \ref{Sec6}  we treat in some detail the $\Nc=2$ case, 
also explaining how to
use the superlocalization formula in the presence of constraints. 

One can notice that these results can in fact
be applied to any supersymmetric theory, whose Lagrangian can be
written as the BRST variation of some function, in any number of
dimensions.

\section{Equivariant cohomology\label{Sec2}}
We recall the basic formulas in equivariant cohomology.
Let $M$ an $n$-dimensional differentiable  manifold,
and $G$ a Lie group with Lie algebra $\g$ and
an action $\rho$ on $M$.
To any $\xi\in\g$ one associates the fundamental
vector field (on $M$)
$$\xi^\ast =\left[\frac{d}{dt}\rho_{(-t\exp\xi)}\right]_{t=0}$$
which we shall write in local coordinates as
$$\xi^\ast=
\xi^\alpha \,T_{\alpha}^i\frac{\partial}{\partial x^i}\,.$$
If $\C[\g]$ is the algebra of polynomial $\C$-valued
functions on $\g$, and $\Omega(M)$ is the
algebra of differential forms on $M$,
the product algebra $\Omega(M,\g)=\C[\g]\otimes\Omega(M)$
has a natural grading, defined by
$$
\deg(P\otimes\beta)=2 \deg(P)+\deg(\beta)
$$
if $P$ and $\beta$ are homogeneous. The group $G$
acts on  $\Omega(M,\g)$ as
\beq\label{action}
(g\cdot\alpha)(\xi)={\rho_g}^\ast(\alpha(Ad_{g^{-1}}\xi)).
\eeq
Moreover one defines the \emph{equivariant differential}
$${d_\g}\colon \Omega(M,\g)^\bullet\to \Omega(M,\g)^{\bullet+1}\,,\qquad(d_\g\alpha) (\xi) =
d(\alpha(\xi))-i_{\xi^*}\alpha(\xi);$$ one has
\beq\label{d2}
({d_\g}^2\alpha)(\xi)=-\Lie_{\xi^\ast}(\alpha(\xi))
\eeq
where $\Lie$ denotes the Lie derivative.
The elements of the invariant subalgebra
$\Omega_G(M)=(\Omega(M,\g))^G$
are called \emph{equivariant forms}; in view of
eq.~(\ref{d2}), on them the
equivariant differential squares to zero,
allowing the definition of the \emph{equivariant cohomology}
of $M$ as $H^\bullet(M,G)=H^\bullet(\Omega_G(M),d_\g)$.

If $M$ is compact, and $\alpha$
is an equivariantly closed element in $\Omega_G(M)$,
and $\xi\in\g$, one denotes by $\int_M\alpha(\xi)$
the integral of the piece of $\alpha(\xi)$ which
is an $n$-form ($n=\dim M$), in the usual gradation of
$\Omega(M)$. This integral can be nicely evaluated
by using a \emph{localization formula}. For every $\xi\in\g$
let $M_\xi\subset M$ be the zero-set of $\xi^\ast$.
If $G$ is compact, and the zeroes of $\xi^\ast$ are isolated
(so that $M_\xi$ is finite), one has
\beq\label{usual}
\int_M \alpha(\xi)=(-2\pi)^{n/2}\sum_{p\in M_\xi}
\frac{\alpha(\xi)_0(p)}{\det^{1/2} L_{p,\xi}}
\eeq
where $\alpha(\xi)_0$ is the 0-form part of $\alpha(\xi)$,
and $L_{p,\xi}\colon T_pM\to T_pM$ is the linear operator defined
as
$$ L_{p,\xi}(v)=[\xi^\ast,v]$$
(this is well defined since $\xi^\ast(p)=0$).

\section{Supersymmetry and vector bundles over the
instanton moduli space\label{Sec3}}
If $M$ is a K\"ahler (or hyperk\"ahler) surface,
and $\Mc$ is the moduli space of instantons on $M$,
via index-theoretic constructions one can define
some vector bundles on $\Mc$. In this section we briefly
review the construction of these bundles and recall
how they are involved in the study of SYM
with $\Nc=1,2$ or $4$ supersymmetries.

We start by considering a K\"ahler surface $M$.
If $E$ is an $SU(N)$ vector bundle on $M$, with $c_2(E)=k$,
there is a smooth space $\Mc$ parametrizing equivalence
classes of irreducible anti-self-dual connections on $E$ (instantons)
modulo gauge equivalence (some subtleties are involved here,
e.g.~the fact that the K\"ahler metric in $M$ should be generic
in a suitable sense, but these issues will be ignored here.
A good reference on this topic is \cite{DK}.)

The choice of a spin structure in $M$ amounts to a choice of a square root
$L$ of the canonical bundle $K$, i.e., a complex line bundle
$L$ such that $L^2\simeq K$ (the canonical bundle $K$ is
the bundle of holomorphic 2-forms on $M$). We shall consider
$L$ as a smooth bundle (i.e., we also allow for smooth sections
rather than just holomorphic). The spin bundles
$S_+$, $S_-$ (i.e., the rank-two complex bundles of undotted and dotted spinors, respectively)
may be taken as
$$S_- = [\Omega^{0,0}\oplus\Omega^{0,2}]\otimes L\,,
\qquad S_+ = \Omega^{0,1}\otimes L\,.$$
It this not difficult to show that $\Omega^{0,2}\otimes L\simeq L^\ast$,
so that $S_-$ splits as a direct sum of orthogonal (in the K\"ahler metric)
subbundles of rank 1, $S_-\simeq L\oplus L^\ast$.

One trivially has $ S_+\otimes L^\ast\simeq  \Omega^{0,1}$; moreover,
since $K\simeq \det \Omega^{1,0}$ if we consider smooth bundles,
one has  $\Omega^{0,1}\otimes K\simeq \Omega^{1,0}$, hence
$S_+\otimes L \simeq \Omega^{1,0}$. Let us describe these isomorphisms
in component notation. Let $(e^0,\dots,e^3)$ be a local orthornomal
frame of 1-forms, and consider the associated local bases of sections
of the spin bundles $S_+$, $S_-$. One has an isomorphism
$S_+\otimes S_- \simeq T^\ast M\otimes \C$, which is given by
$$\psi\otimes\chi \mapsto \sigma_{m\alpha\dot\beta}\,\psi^\alpha\,\chi^{\dot\beta}
\,e^m\,,$$
where $\sigma_{m\alpha\dot\beta}$ are the Pauli matrices. Thus
 the isomorphisms
$S_+\otimes L \simeq \Omega^{1,0}$, $ S_+\otimes L^\ast  \simeq\Omega^{0,1}$
express the fact that the forms
$$\sigma_{m\alpha \dot 1}\,e^m\,,\qquad \sigma_{m\alpha \dot 2}\,e^m$$
are of type (1,0) and (0,1), respectively. All this is explained
in more ``physical'' terms in \cite{Witten95,J}.

{\bf $\Nc=2$ supersymmetry.}
Let $E$ be an $SU(N)$ bundle on $M$, and let $\Mc$
be the moduli space of instantons on $E$. For every $m\in\Mc$ we have a twisted
Dirac operator $$D_m\colon S_+\otimes S_- \otimes \mbox{ad}(E)\to S_-\otimes S_-\otimes  \mbox{ad}(E)$$
(where $\mbox{ad}(E)$ is the adjoint bundle of $E$, i.e., the bundle of
trace-free endomorphisms of $E$ and the global symmetry group of the $\Nc=2$ theory is identified with
the $SU(2)$ structure group $S_-$),
together with the adjoint operator  $D_m^\ast$. The assignment
$$ m \in\Mc\quad\rightsquigarrow\quad\ker D_m -\mbox{coker}\, D_m=\ker D_m-\ker D_m^\ast$$
defines the index $\mbox{ind}(D)$ as a class in the topological K-theory
of $\Mc$. Under our assumptions (in particular, we consider only irreducible
instantons) we have $\ker D_m^\ast =0$ for all $m\in \Mc$, so that $\mbox{ind}(D)$
is a vector bundle, that we denote by $\mathcal S_2$. The fibre of $\mathcal S_2$
at the point $m\in\Mc$ is the vector space $\ker D_m$. This (complex) vector bundle
may be naturally identified with the complexified tangent bundle to $\Mc$ \cite{AHS},
and one has $\mbox{rk}\, \mathcal S_2 = \dim_\R \Mc$ (which
equals $4kn$ in the case of framed $SU(N)$ instantons on
$\R^4$). Since we have considered
the full spin bundle $S_+$, this is the case relevant to $\Nc=2$
supersymmetry.

{\bf $\Nc=1$ supersymmetry.} In the case $\Nc=1$ one  considers the
subbundle $L$ instead of the full bundle $S_+$.  In this way we get
a vector bundle $\mathcal S_1$ on $\Mc$ with $\mbox{rk}\, \mathcal S_1 = \frac12\,\dim \Mc$.
One can notice that the orthogonal splitting $S_-=L\oplus L^\ast$ implies
that the complexified tangent bundle to the instanton moduli space
$T\Mc\otimes\C\simeq\mathcal S_2$
splits as a direct sum of two vector bundles,
$T\Mc\otimes\C\simeq\mathcal S_1\oplus\mathcal S'_1$,
which are dual to the bundles of
differential forms of type (1,0) and (0,1) on $\Mc$. One has $\mbox{rk}\, \mathcal S_1
= 2kN$ for the framed instantons.

{\bf $\Nc=4$ supersymmetry.} In this case we have a family of Dirac operators
$$D\colon S_+ \otimes \Sigma \otimes  \mbox{ad}(E) \to
S_- \otimes \Sigma \otimes  \mbox{ad}(E)$$
where $\Sigma$ is a bundle with structure group
$SU(4)$.\footnote{However as a consequence of the twisting of the theory
\cite{VW,M,Y} the structure group is reduced to a little group.
Of the three possibilities listed in  \cite{VW,M,Y} we discuss
in \cite{BFMT} the
so-called $SU(2)\times SU(2)\times U(1)$ twisting, where the
structure group is actually reduced to $S(U(2)\times U(2))\subset
SU(4)$.}
For framed instantons the resulting bundle on $\Mc$ has rank
$8kN$.

\section{Supermanifolds\label{Sec4}}
We want to introduce some supermanifolds whose ``bosonic'' part
is the instanton moduli space $\Mc$; this will be done
by using the vector bundles on $\Mc$ that we have introduced in the previous
section. We start by briefly recalling the basic definitions about
supermanifolds. (We consider supermanifolds in the sense of
Berezin-Le\u \i tes and Konstant \cite{BL,K}).

An $(m,n)$ dimensional supermanifold $\Mg$ is a pair $(M,\A)$, where $M$ is an
$m$-dimensional manifold, and $\A$ is a sheaf of $\Z_2$-graded commutative
algebras on $M$, such that:

\begin{enumerate}  \item there is morphism of sheaves of $\R$-algebras
$\varepsilon\colon \A\to\mathcal C_M^\infty$;
\item if $\Nc$ is the nilpotent subsheaf of $\A$, the quotient
 $\Nc/\Nc^2$ is a locally free sheaf (vector bundle)
$E  $ of rank $n$;
\item $\A$ is locally isomorphic to the exterior algebra sheaf
$\Lambda^\bullet \E  $, and this
isomorphism is compatible with the morphism $\varepsilon$.
\end{enumerate}

These conditions imply $\ker\varepsilon=\Nc$. Also, condition
3 implies that $\varepsilon$ is surjective (as a
sheaf morphism).

If we start from a rank $n$ vector bundle $E  $ on $M$
we may construct an $(m,n)$-dimensional supermanifold $(M,\A)$
by letting $\A=\Lambda^\bullet\E  $, with $\varepsilon$ the natural
projection $\Lambda^\bullet E  \to\E  $.
Conversely, standard arguments in deformation theory allow one
to prove that any supermanifold is \emph{globally} isomorphic
to a supermanifold of this kind \cite{Batch}.

If $(x^1,\dots,x^m)$ are local coordinates in
$M$, and $(\theta^1,\dots,\theta^n)$ is a local basis of sections of
$\E  $,
then the collection
$(x^1,\dots,x^m,\theta^1,\dots,\theta^n)$
is   a local coordinate chart for $\Mg$.
According to the  requirements above,
a   section of $\A$ (i.e., a superfunction on $\Mg$) has a local expression
\begin{equation}
f = \sum_{k=1}^n f_{[k]} \label{superfield}
\end{equation}
where
$$f_{[k]} = \sum_{\alpha_1\dots\alpha_k=1\dots n} 
f_{\alpha_1\dots\alpha_k}(x)\,\theta^{\alpha_1}\cdots\theta^{\alpha_k}$$
with $f_0=\varepsilon(f)$. Thus we get the well-known   superfield expansion.

If $\A_p$ is the stalk of $\A$ at $p\in M$ (i.e., the algebra
of germs of sections of $\A$ at $p$), then
$$\mathcal I_p=\ker \varepsilon\colon \A_p \to (\Cc^\infty_M)_p$$
is a maximal graded  ideal of $\A_p$.
The \emph{tangent superbundle} is by definition the sheaf
$\mathcal Der_\R\A$ of graded derivations of $\A$. The tangent
superspace $T_p\Mg$ at a point $p\in M$ is by definition the
$(m,n)$-dimensional graded vector space
$$T_p\Mg = \mathcal Der_\R\A/\mathcal I_p\cdot \mathcal Der_\R\A$$
and one has a canonical isomorphism
$T_p\Mg \simeq T_pM \oplus \E_p^\ast$
where $\E^\ast_p$ is the fibre at $p$ of the dual vector bundle
$\E^\ast$, or, in terms of superbundles,
$$ T\Mg\simeq \A\otimes[TM\oplus\E^\ast ]\,.$$

{\bf Tautological supermanifolds.} Our strategy for the
computation of the partition function of the topological SYM
theory will involve considering supermanifolds based
on the instanton moduli space $\Mc$, associated with the vector
bundles $\mathcal S_1$, $\mathcal S_2$, $\mathcal S_4$ we
have introduced in the previous section. Since we shall study
in detail the case of $\mathcal N=2$ supersymmetry, and since
$\mathcal S_2$ may be identified with the tangent bundle
to $\Mc$, we shall develop in some detail the case where
the vector bundle to which a supermanifold is associated with
is the cotangent bundle (we choose the cotangent rather than the
tangent bundle for mere reasons of convenience).

So, if given a supermanifold $\Mg=(M,\A)$, if the associated
vector bundle $\E  $ on $M$ is isomorphic
to the cotangent bundle $T^\ast M$, we say
that $\Mg$ is \emph{tautologically associated with $M$}.
If the isomorphism $\A\simeq\Lambda^\bullet T^\ast M$ has been fixed, one has
a canonical isomorphism
\beq\label{supertang} T\Mg\simeq \A\otimes[TM\oplus TM]\eeq
(cf. \cite{BFTT}), and there is a naturally defined involution $\Pi$
on $T\Mg$, exchanging the two summands.

Superfunctions on $\Mg$ are just differential forms on $M$; one has
an isomorphism (of sheaves of $\Cc_M^\infty$-modules)
 $\tau\colon\Omega^\bullet\to\A$. Note that if $(x^1,\dots,x^m)$
are local coordinates in $M$, and we set $\theta^i=\tau(dx^i)$,
then $(x^1,\dots,x^m,\theta^1,\dots,\theta^m)$ is a local
coordinate system for $\Mg$ (and of course, $d\theta^i\ne0$!).

{\bf Berezin integration.}\footnote{This approach to the Berezin integral
is taken from \cite{HM}.}
Let $\Mg=(M,\A)$ be an $(m,n)$ dimensional supermanifold, with $M$  an
oriented manifold, and denote
by $\Omega^m_\Mg$ the sheaf of super differential $m$-forms on $\Mg$, and by
$\mathcal P_n$ the sheaf of graded differential operators of order $n$ on
$\A$. The sheaf   $\Omega^m_\Mg$ has its natural structure of graded left
$\A$-module given
by multiplication of forms by functions; the sheaf $\mathcal P_n$ has an
an analogous graded left $\A$-module structure, but   also has a (inequivalent)
right $\A$-module structure, given by
$$(D\cdot f)(g)=D(fg)$$ where $f$, $g$ are superfunctions. We consider on
 $\mathcal P_n$ this module structure, and take the graded tensor product
$\Omega_\Mg^m\otimes_\A\mathcal P_n$.

The   structural morphism $\varepsilon\colon\A\to\mathcal C_M^\infty$
extends to a morphism $\Omega^m_\Mg\to\Omega_M^m$, whose action we denote by
 a tilde.
The sheaf $\Omega_\Mg^m\otimes_\A\mathcal P_n$ has a subsheaf
$\mathcal K$
whose sections $\omega$ are such that the differential $m$-form
$\widetilde{\omega(f)}$
on $M$ is exact for every choice of a superfunction $f$ with compact
 support (more precisely,
$\widetilde{\omega(f)}=d\eta$ for a compactly supported $(m-1)$-form
$\eta$ on $M$).
The quotient $\Omega_\Mg^m\otimes_\A\mathcal P_n/\mathcal K$ is denoted
by $\ber(\Mg)$
and is called the \emph{Berezinian sheaf} of $\Mg$. It is a locally free
graded $\A$-module
of rank (1,0) if $n$ is even, rank (0,1) is $n$ is odd. On its
compactly supported sections one can define an integral
(the Berezin integral) by letting
$$\int_\Mg \omega = \int_M \widetilde{\lambda(1)}$$
where $\lambda$ is any section of $\Omega_\Mg^m\otimes_\A\mathcal P_n$
whose class in the quotient   $\Omega_\Mg^m\otimes_\A\mathcal P_n/\mathcal K$
is $[\lambda]=\omega$.
This integral performs the usual procedure of ``integrating over
the fermions'': indeed, given a local coordinate
system   $(x^1,\dots,x^m,\theta^1,\dots,\theta^n)$ defined in a patch $U$,
one has
$$\omega_{\vert U} =
\left[dx^1\wedge\dots\wedge dx^m\otimes \vf{}
{\theta^1}\dots\vf{}{\theta^n}\right]f$$
for a superfunction $f\in\A(U)$,
and if $\omega$ is supported in $U$, one has
$$\int_\Mg \omega = \int_M f_{[ n]}\,dx^1\dots dx^m$$
i.e. the Berezin integral is the usual integral over $M$ of the last term
in the superfield expansion eq.~(\ref{superfield}) of the component
superfunction $f$, and corresponds to the usual operation of
``integrating over the fermions'' in quantum field theory.

The Berezinian bundle of a `tautological' supermanifold $\Mg$ has
a canonical global section  $\Theta$. If $\Theta_0$ is
a global nowhere vanishing differential $n$-form on $M$,
and $\Delta$ is a dual derivation of order $n$ (i.e., $\Delta(\Theta_0)=1$),
the class  $\Theta=[\Theta_0\otimes\Delta]$ is a well-defined
global section of $\ber(\Mg)$, independent of the choices
of $\Theta_0$ and $\Delta$. In local coordinates
$(x,\theta)$, where $\theta^i=dx^i$, one has
$$\Theta=\left[dx^1\wedge\dots\wedge dx^n\otimes
\frac{\partial}{\partial \theta^1}\dots\frac{\partial}
{\partial \theta^n}\right]\,.$$
If $\eta$ is an $n$-form  on $M$, one has
$$\int_{\Mg}\Theta\,\tau(\eta) = \int_M\eta\,.$$

\section{BRST transformations and superlocalization formulas\label{Sec5}}
Let  $\Mg=(M,\A)$ be  an
$(m,n)$-dimensional supermanifold, with $\A$ the sheaf of
sections of the exterior algebra bundle of  a rank $n$ vector bundle
$\E  $ on $M$. Assume that there is an action $\rho$ of a Lie group $G$ on
$M$, and that $G$ also acts on $\E  $ by a linear action $\hat\rho$
in such a way that the diagram
$$\xymatrix{\E   \ar[d] \ar[r]^{\hat\rho_g} & \E   \ar[d] \\
M \ar[r]^{\rho_g} & M}$$
commutes for all $g\in G$. The action $\hat\rho$ induces
a vector field $\hat\xi^\ast$ on $\E$:
$$\hat\xi^\ast= \left[\frac{d}{dt}\hat\rho_{\exp(- t\xi)}\right]_{t=0}\,.$$
If $(x^1,\dots,x^m)$ are local coordinates on $M$, and
$(\theta^1,\dots,\theta^n)$ are a local basis of sections of $\E$
(so that they can be regarded as local fibre coordinates on $\E^\ast$),
$\hat\xi^\ast$ is locally written as
\beq\hat\xi^\ast =  \xi^\alpha \,T_\alpha^i\,\vf{}{x^i} +
\xi^\alpha \, \theta^B\, U_{\alpha B}^A\,\frac{\partial}{\partial \theta^A}\label{classrecipe}\eeq
where the functions  $\xi^\alpha \,T_\alpha^i$ are the local components
of the generator $\xi^\ast$ of the action of $G$ on $M$. The vector field
(\ref{classrecipe}) can be regarded as an even  super vector field $\hat\xi^\ast$
on $\Mg$ representing the induced  infinitesimal action of $G$ on $\Mg$.

If $p\in M$ is a zero of $\xi^\ast$, then an endomorphism $\tilde L_{\xi,p}$ of
the fibre $\E_p$ rests defined. We may regard this as an even endomorphism
${\bf L}_{\xi,p}$ of the cotangent superspace $T_p^\ast\Mg$, defined as
$(1,L)$ after identifying $T_p^\ast\Mg\simeq T_p^\ast M\oplus \E_p$.

We shall consider the algebra $\C[\g]\otimes\A(M)$ (where
$\A(M)$ is the algebra of global sections of $\A$), which carries
the action of $G$ given by 
$$(g\cdot\alpha)(\xi) = \hat\rho_g(\alpha(\mbox{Ad}_{g^{-1}}\xi))$$
where by abuse of notation we denote by $\hat\rho$ the induced action
of $G$ on $\Mg$.  On $\C[\g]\otimes\A(M)$ one considers the $\Z$-grading
$$\deg(P\otimes f)=2\deg(P) + \deg(f)$$
(where $\deg(P)$ is the degree of the polynomial $P\in \C[\g]$
and $\deg(f)=k$ if $f=f_{[k]}$) and the $\Z_2$-grading
given by the grading of $\A$ (the terms ``even'' and ``odd'') will refer to this grading). We shall denote by $\Ag_G$ the subalgebra
of $\C[\g]\otimes\A(M)$ formed by $G$-invariant elements.

\begin{defin} \label{defBRST} A BRST operator is an odd derivation $Q$ of $\C[\g]\otimes\A(M)$ of $\Z$-degree 1 such that 

1. $(Q^2 F)(\xi)=\hat\xi^\ast(F)$ for all $F\in \C[\g]\otimes\A(M)$ and $\xi\in
\g$;

2. $Q$ is equivariant, i.e., $Q\circ g=g\circ Q$ for all $g\in G$;

3. The equivariant morphism $\sigma_Q\colon\E^\ast\to TM$ defined by
$$ \sigma_Q(v)(f) = i_v(Q(f))\qquad\mbox{for all functions $f$ on $M$}$$
is injective.
\end{defin}

\begin{remark} \rm The third condition fails in the case
relevant to $\Nc=4$ supersymmetry, and should rather be replaced
by the condition that $\E$ is a direct sum in such a way that
$\sigma_Q$ is injective after restriction to any of the summands
of $\E^\ast$. However for the sake of simplicity we shall stick
to this condition in its present form.
\end{remark}

In particular, this implies that $Q^2_{\vert\Ag_G}=0$, so that 
an equivariant cohomology $H^\bullet(\Ag_G,Q)$ is defined.
Moreover, for every $\xi\in\g$ one can define an odd supervector field
$Q_\xi$ by letting $Q_\xi(F(\xi))=Q(F)(\xi)$ for all $F\in \C[\g]\otimes\A(M)$.
In terms of this supervector field the first requirement in Definition \ref{defBRST}
reads
\beq\label{compare} [Q_\xi,Q_\xi] = 2 \hat\xi^\ast \eeq
where $[\,,\,]$ is the graded commutator of supervector fields (in this
case an anticommutator in fact). Let us, for future use, write this equation
in local components. After writing
$$Q_\xi = a^i\,\frac{\partial}{\partial x^i} + b^A\,\frac{\partial}{\partial \theta^A}\,,$$
eq.~(\ref{compare}) is equivalent to the conditions
\beq\label{compofQ} 
a^i = \sigma^i_A\,\theta^A\,,\qquad b^A\,\sigma^i_A= \xi^\alpha\,T_\alpha^i\eeq
where $\sigma^i_A$ is the matrix representing the morphism
$\sigma_Q^\ast\colon T^\ast M\to \E$, i.e., $\sigma_Q^\ast(dx^i)=\sigma^i_A\,\theta^A$.

\begin{lemma} If $p\in M$ is a zero of $\xi^\ast$ for an element $\xi\in\g$, 
the diagram
\beq\label{ends}\xymatrix{T_p^\ast M \ar[r]^{\sigma_Q^\ast} \ar[d]_{\mathcal L_{p,\xi}}& 
\E_p \ar[d]^{{\tilde L}_{\xi,p}} \\ T_p^\ast M \ar[r]^{\sigma_Q^\ast} &\E_p}\eeq
commutes.
\end{lemma}
\begin{proof} The commutativity of the diagram is equivalent
to the infinitesimal equivariance of the morphism $\sigma_Q$.
\end{proof}

Finally, we require the existence of a  $G$-invariant Riemannian metric $h$ on $M$.
Since the morphism $\sigma_Q$ is injective this also defines a  $G$-invariant
fibre metric $H$ on $\E$ by letting  $H(u,v)=h(\sigma_Q(u),\sigma_Q(v))$ for all $u,v\in\E^\ast$.
Using the metrics $h$ and $H$ one can construct a $G$-invariant global
section of the Berezinian sheaf $\mbox{Ber}(\Mg)$, whose local
coordinate expression is
$$\Theta = \left[dx^1\wedge\dots\wedge dx^m\otimes
\frac{\partial}{\partial \theta^1}\dots\frac{\partial}
{\partial \theta^n}\right]\,\frac{\det ^{1/2}(h)} {\mbox{det}^{1/2}(H)}
\,.$$
Actually, this Berezinian measure does not depend on the metrics but only on the BRST operator 
$Q$ via the morphism $\sigma_Q$.

\begin{assumption} The morphism $\sigma_Q$, regarded as a section
of the bundle $\E\otimes TM$, is parallel with respect to the connection
induced by the metrics $h$ and $H$.
\label{assump}\end{assumption}

\begin{lemma}\label{superstokes} Let $\nu$ be a superfunction which is homogeneous
of degree $n-1$, i.e., $\nu=\nu_{[n-1]}$, and let $\ast_H\nu$ be the
section of $\E^\ast$ which is Hodge dual to $\nu$ via the metric $H$,
i.e., in local coordinates,
$$(\ast_H\nu)^A=\frac{1}{(n-1)!}(\det (H))^{-1/2}\,\varepsilon^{AA_2\dots A_n}\,
\nu_{A_2\dots A_n}\qquad\mbox{if}\qquad \nu =\nu_{A_1\dots A_{n-1}}\,\theta^{A_1}\dots
\theta^{A_{n-1}}\,,$$
where $\varepsilon^{A_1\dots A_n}$ is the completely antisymmetric symbol. Then,
for every regular domain $U\subset M$ with compact closure,
$$\int_{\Mg\vert U }\Theta \,Q_\xi(\nu) = \int_{\partial U} \ast_h\sigma_Q(\ast_H\nu))\,;$$
here $\partial U$ is equipped with the induced orientation, and $\ast_h$ is Hodge
duality in $M$.
\end{lemma}
\begin{proof} The   equality is proved by direct computation. It is necessary to use
the Assumption \ref{assump}.
\end{proof}

We can now state the localization formula. Let $Q$ be a BRST operator.

\begin{thm} Let $M$ and $G$ be compact, let $F\in\Ag_G$ be such that $Q(F)=0$, and assume that $\xi\in \g$ is such that $\xi^\ast$ only has isolated zeroes. Then,
\beq\label{superlocform}\int_{\Mg}\Theta\,F(\xi) = \frac{(-2)^{n/2}(n/2)!\pi^{m/2}}{(m/2)!} \sum_{p\in M_\xi}
\mbox{\rm Sdet}^{1/2}({\bf L}_{p,\xi})\, F(\xi)_0(p)\eeq
where $\mbox{\rm Sdet}({\bf L}_{p,\xi})$ is the superdeterminant (Berezinian determinant) of the even endomorphism ${\bf L}_{p,\xi}$ (cf.~\cite{B,K,BHR}).
\label{superloc}
\end{thm}
The proof of this localization formula follows the pattern of the
proof of the usual formula, cf.~\cite{BGV}. So we need the following
preliminary results. We assume that $M$ and $G$ are both compact,
and an element $\xi\in\g$ such that $\xi^\ast$ has isolated zeroes
has been fixed.
\begin{lemma}  There exists a superfunction  $\beta$
(actually, a section of $\E$)
such that 

1. $\hat\xi^\ast(\beta)=0$;

2. $Q_\xi(\beta)$ is invertible outside $M_\xi$;

3. Every $p\in M_\xi$ has a neighbourhood on which
the function $H(\beta,\beta)$  equals the square distance from the point $p$.
\end{lemma}
\begin{proof} One can construct a differential 1-form $\lambda$ on $M$
such that $\Lie_{\xi^\ast}(\lambda)=0$ and $\lambda(\xi^\ast)=d_p^2$,
where $d_p(x)$ is the distance of $x$ from $p$ in the metric $h$
\cite{BGV}. The section $\beta=\sigma_Q^\ast(\lambda)$
of $\E$ satisfies the required conditions. 
\end{proof}

\begin{lemma} The superfunction $F(\xi)_{[n]}$ is Q-exact outside $M_\xi$, i.e., there
is a section $\nu$ of $\A$ on $M\setminus M_\xi$ such that
$$ {F(\xi)_{[n]}}_{\vert M\setminus M_\xi} = Q_\xi (\nu)\,.$$
\end{lemma}
\begin{proof} In view of the previous Lemma, outside $M_\xi$ we may set
$$\nu = \left(\beta\,F(\xi)\,Q_\xi(\beta)^{-1}\right)_{[n-1]}\,,$$
and again using the previous Lemma, one gets the desired equality.
\end{proof}

\noindent \emph{Proof of Theorem \ref{superloc}.} For every $p\in M_\xi$
let $B_\epsilon(p)$ be the ball of radius $\epsilon$ (measured with the metric $h$)
around $p$, and denote by $\Mg_\epsilon$ the supermanifold $\Mg$
restricted to the complement of the union of the closures of the balls $B_\epsilon(p)$. Then,
using Lemma \ref{superstokes},
$$ \int_{\Mg}\Theta\,F(\xi) = 
\lim_{\epsilon\to 0} \int_{\Mg_\epsilon}\Theta\,
Q_\xi(\nu) = -  \lim_{\epsilon\to 0}
\sum_{p\in M_\xi} \int_{S_\epsilon(p)}\ast_h\sigma_Q(\ast_H\nu)$$
where $S_\epsilon(p)$ is the boundary of $B_\epsilon(p)$.
Under the rescaling $$x\mapsto \epsilon^{1/2}\,x,\qquad
\theta\mapsto \epsilon^{1/2}\,\theta$$
the term $\mu=\beta\,Q_\xi(\beta)^{-1}$ is homogeneous of degree zero,
so that we get 
$$  \int_{\Mg}\Theta\,F(\xi) =  - \sum_{p\in M_\xi}  F(\xi)_0(p) \int_{S_1(p)}\ast_h \sigma_Q(\ast_H\mu_{[n-1]})\,.$$
The integrals in the r.h.s., again using Lemma \ref{superstokes},
may be recast as Berezin integrals over the supermanifolds 
$\Mg\vert B_1(p)$. These may be evaluated by writing their
 integrands in local coordinates, obtaining  
$$(-1)^{n/2} \int_{\Mg\vert B_1(p)} \Theta\,
\left[ \sum_{A,B=1}^n a_{AB}(p)\,\theta^A\,\theta^B\right]^{n/2}$$
where $a_{AB}(p)$ is a skew-symmetric matrix of constants 
which represents the morphism $\tilde L_{p,\xi}$ (with an index lowered
with the metric $H$). From this we get (cf.~e.g. \cite{ZJ})
$$\int_{\Mg}\Theta\,F(\xi) =(-2)^{n/2}(n/2)! \sum_{p\in M_\xi}  F(\xi)_0(p) \,\mbox{Pf}(a(p))\,\int_{B_1(p)} \mbox{vol}(h)$$
where $\mbox{vol}(h)$ is the Riemannian volume form, and $\mbox{Pf}(a(p))$
is the Pfaffian of the matrix $a(p)$.
Since
$$\mbox{Pf}(a(p)) = \mbox{\rm Sdet}^{1/2}({\bf L}_{p,\xi})\qquad\mbox{and}\qquad
\int_{B_1(p)} \mbox{vol}(h)=\frac{\pi^{m/2}}{(m/2)!}$$
we eventually obtain the superlocalization formula.

\begin{example}\rm The simplest example is provided by the tautological
supermanifolds; then $\E=T^\ast M$ and for every $\xi\in \g$ the vector
field $Q_\xi$ is
$$Q_\xi = d + \Pi(\xi^\ast,0)$$
where the exterior differential $d$ is regarded as an odd
supervector field on $\Ag$, and $\Pi$ is the morphism which interchanges
the two summands in eq. (\ref{supertang}).
$\sigma$ turns out to be   the identity
morphism. The superlocalization formula reduces
to the usual localization formula (eq.~(\ref{usual})); note that
$\mbox{\rm Sdet}^{1/2}({\bf L}_{p,\xi})=\mbox{det}^{-1/2}(L_{p,\xi})$. In this case the
isomorphism $\tau\colon\Omega^\bullet\to \A$ intertwines the
equivariant differential with the BRST operator $Q$.
This is the suitable framework for  $\Nc=2$
supersymmetry.
\end{example}

\begin{example}\rm Let $M$ be a K\"ahler manifold of even complex dimension $m$,
with K\"ahler form
$\omega$, and assume that $G$ acts on $M$ by K\"ahler isometries
(so every $\rho_g$ is a holomorphic isometry for the K\"ahler metric
$h$). After complexifying the tangent bundle $TM$, we take $\E=\Omega^{1,0}$,
and $$Q_\xi =\partial + \Pi(\xi^\ast,0)\,.$$
The morphism $\sigma^\ast_Q$ is the projection $T^\ast M\otimes\C\to\Omega^{1,0}$.
We obtain the superlocalization formula eq. (\ref{superlocform}) with numerical factor
$(-2)^{m/2}\pi^m(m/2)! / m!$. This is the picture relevant to $\Nc=1$
supersymmetry. \end{example}

\section{Application to topological $\Nc=2$ SYM\label{Sec6}}
We want to apply the superlocalization formula
to the computation of the partition function
for topological Yang-Mills theory. We consider explicitely the case
of $\Nc=2$ supersymmetry but the cases $\Nc=1,4$
may be dealt with along the same lines after choosing the
relevant supermanifolds on the instanton moduli space.

We start by
briefly recapping the ADHM construction for framed $SU(N)$
instantons on $\R^4$.\footnote{The original source
is \cite{ADHM};  a useful reference on this construction,
and other issues that will be touched upon in this and the following
sections, is \cite{Naka}}  Framed instantons are  anti-self-dual $SU(N)$
connections on (trivial bundle on) $\R^4$ with a fixed
framing at infinity (i.e., if we transfer the instanton to
the sphere $S^4$ via a stereographic projection, there
is a fixed isomorphism of the fibre at a given point with $\C^N$,
and this isomorphism is part of the data specifying the
instanton). The  moduli space of framed instantons under gauge
 equivalence is a
singular manifold of dimension $4kN$, where  $k$ in the
second Chern class (instanton number) of the instanton.
The ADHM description is obtained in terms of data
consisting of $k\times k$ matrices $B_1$, $B_2$,
a $N\times k$ matrix $I$ and a $k\times N$ matrix $J$,
all with complex entries. These are subject to the constraints
\beq\label{c1}
[B_1,B_1^\dagger]+[B_2,B_2^\dagger]+II^\dagger-J^\dagger J=0\eeq
\beq\label{c2}
[B_1,B_2]+IJ=0
\eeq
where $^\dagger$ denotes hermitian conjugation. The group $U(k)$
acts on these data, by adjunction on $B_1$, $B_2$ and by multiplication
from the suitable side on $I$ and $J$, and this action
preserves the constraints.
The moduli space of framed instantons is obtained
by taking all the data $(B_1,B_2,I,J)$ satisfying the
constraints and taking equivalence classes under the action
of $U(k)$. The resulting moduli space is singular, its smooth points
corresponding to data with trivial stabilizer under the
$U(k)$ action. Singularities may be resolved, either
by standard blowup techniques, or  using the hyperk\"ahler
quotient costruction of the moduli space. We
shall denote by $\Mc$ the smooth moduli space
so obtained.

In SYM one supplements the ADHM data
by fermionic moduli provided by the zero modes of
the gaugino field.  For $\Nc=2$ the
fermionic moduli can be identified with
differential  forms on the bosonic moduli space $\Mc$;
this is the reason of the introduction of the ``tautological
supermanifolds'' of Section \ref{Sec4}.
The constraints on the fermionic data are obtained
by linearizing the bosonic constraints, and the multi-instanton action
eventually obtained is
obtained by plugging into the SYM action the bosonic and fermionic
zero modes in terms of the (unconstrained) ADHM data and imposing the ADHM
constraints via Lagrangian multipliers.
The resulting action turns out to be  BRST-exact,
hence, given its invariance under action of the
groups involved, also BRST-closed.

Moreover as we have hinted in the previous Sections,
if one associates a tautological supermanifold to the
bosonic moduli space, the operator $Q$ of the previous
section --- the counterpart on the supermanifold side of
the equivariant differential --- is exactly the BRST operator.
Putting all this together, this opens the way
to the computation of integrals over the moduli space
of quantities depending on the SYM action, such as the
partition and correlation functions, by means
of a superlocalization formula.

We shall at first ignore the existence of the
constraints on the ADHM data.
So the field content of the theory is given by
the matrices $B_1$, $B_2$, $I$, $J$ with their fermionic partners
$M_1$, $M_2$, $\mu_I$, $\mu_J$.
We would like to consider the action of the group
$U(k)\times SU(N)$, but if we do so the fixed points of the group
action will not be isolated. It is therefore convenient,
following Nakajima \cite{Naka}, to introduce also
an action of the group  $T^2$, given by
$$ (B_1,B_2,I,J)\mapsto (e^{i\epsilon_1}B_1,
e^{i\epsilon_2}B_2, I,\
e^{i(\epsilon_1+\epsilon_2)}J).
$$
If we denote by $\phi$, $a$, $(\epsilon_1,\epsilon_2)$
elements in the Lie algebras of $U(k)$, $SU(N)$, $T^2$ respectively,
we obtain for the vector fields $\xi^\ast$ and $Q_\xi$ the
following expressions:
\begin{eqnarray*}\xi^\ast &=& (\phi I-Ia) \frac{\partial}{\partial I} +
(-J\phi+aJ+\epsilon J)\frac{\partial}{\partial J}
+ ([\phi,B_\ell]+\epsilon_\ell)\frac{\partial}{\partial B_\ell}\end{eqnarray*}
\begin{eqnarray*}
Q_\xi &=& \mu_I\frac{\partial}{\partial I}  +
\mu_J\frac{\partial}{\partial J} + M_\ell \frac{\partial}{\partial B_\ell}
\\ &+& (\phi I-Ia) \frac{\partial}{\partial \mu_I} +
(-J\phi+aJ+\epsilon J)\frac{\partial}{\partial \mu_J} +
([\phi,M_\ell]+\epsilon_\ell M_\ell)\frac{\partial}{\partial M_\ell}
 \label{BRST2} \end{eqnarray*}
(here $\epsilon=\epsilon_1+\epsilon_2$). One recognizes
in $Q_\xi$ the standard expression of the infinitesimal
BRST transformations in the theory under consideration.

{\bf Introduction of constraints.} If $N\subset M$ is a submanifold of $M$,
locally given in some coordinate patch $(x^1,\dots,x^n)$
by a set of constraints $V_1=\dots=V_r=0$, we may consider the
tautological supermanifold
$\mathfrak N=(N,\mathcal B)$ associated to the cotangent bundle $T^\ast N$.
\begin{prop}\label{supercon} $\mathfrak N$ is a sub-supermanifold of
the tautological supermanifold $\Mg$,
whose  equations in the local coordinate patch
$(x^1,\dots,x^n,\theta^1=dx^1,\dots\theta^n=dx^n)$
are $$V_1=\dots=V_r=0,\qquad W_1=\dots=W_r=0,$$
where the superfunctions $W_i$ expressing the fermionic constraints are given by
$$W_a=\frac{\partial V_a}{\partial x^k}\,\theta^k.$$
\end{prop}
We consider now a situation where the coordinates in $M$ are the bosonic ADHM parameters
which appear in the  Lagrangian $L$ of a $\Nc=2$ SYM \cite{Dorey:1996hu,Fucito:2001ha}
and the functions $V_a$ express 
the ADHM constraints.
One also introduces fermionic partners $\theta$, subject to the constraints
$W_a=0$. The constraints are implemented by the Lagrange multipliers $H^a$ and $\chi^a$, so that
one considers a Lagrangian
$$L'=L+H^a\,V_a(x,\theta)+\chi^a\,\frac{\partial V_a}{\partial x^k}\,\theta^k\,.$$
The Lagrange multipliers should be considered as additional coordinates on an
enlarged supermanifold $\Mg'$.
We have a BRST vector field $Q_\xi$ for the unconstrained theory, and we want to complete it to a new field
$$Q'_\xi = Q_\xi + \tilde Q_\xi= Q_\xi - R^a\,\vf{}{H^a}-S^a\,\vf{}{\chi^a}$$
which leaves the Lagragian $L'$ invariant. To simplify the treatment we assume that
the odd superfunctions $R^a$ are linear in the coordinates $\chi$, i.e.,
$R^b = \chi^a\,N_a^b$ for a matrix of ordinary functions $N$. Since
$$Q'_\xi(L')=Q_\xi(L)-R^a\,V_a+H^a\,\frac{\partial V_a}{\partial x^k}\,\theta^k
-S^a\,\frac{\partial V_a}{\partial x^k}\,\theta^k -\chi^a\, \frac{\partial V_a}{\partial
x^k}\,\xi^\alpha T_\alpha^k$$
and $Q_\xi(L)=0$ we obtain the conditions
\beq\label{newc}S^a=H^a\,,\qquad   N_a^b\,V_b =- \frac{\partial V_a}{\partial
x^k}\,\xi^\alpha T_\alpha^k\,.\eeq
One assumes that the group $G$ acts also on the ``new sector'' of the supermanifold
$\Mg'$, so that the vector field $\hat\xi^\ast$ acquires  a new contribution
$$\tilde\xi^\ast = \xi^\alpha\,\tilde T_\alpha^a\,\vf{}{H^a}
+ \xi^\alpha\,\chi^b\frac{\partial\tilde T_\alpha^a}{\partial H^b}\,\vf{}{\chi^a};$$
one should have $[\tilde Q_\xi,\tilde Q_\xi]_+=2\tilde\xi^\ast$, which is equivalent
to
$$R^a=\xi^\alpha\,\chi^b\,\frac{\partial\tilde T_\alpha^a}{\partial H^b}\,,\qquad
S^a\,\vf{R^b}{\chi^a}=\xi^\alpha\,\tilde T_\alpha^b.$$
If the functions $\tilde T$ are linear in the $H^a$, these conditions are solved
by
\beq\label{addBRST} N^a_b=\xi^\alpha\,\frac{\partial\tilde T_\alpha^a}{\partial H^b}.\eeq
The second equation in eq.~(\ref{newc}) becomes
\beq\label{newc2}\xi^\alpha\,\frac{\partial\tilde T_\alpha^a}{\partial H^b}\,V_b =
- \frac{\partial V_a}{\partial
x^k}\,\xi^\alpha T_\alpha^k\,.\eeq
Provided that this constraint is satisfied,  eq.~(\ref{addBRST})
yields a solution to the problem of finding the BRST transformations
for the Lagrange multipliers.

Since
$$[\vf{}{\theta^k},Q'_\xi]=[\vf{}{\theta^k},Q_\xi] = \vf{}{x^k}\,,\qquad
[\vf{}{\chi^a},Q'_\xi]=[\vf{}{\chi^a},\tilde Q_\xi] = N_a^b\,\vf{}{H^b}$$
the additional requirement for the conditions of Definition \ref{defBRST}
to hold is that the matrix $N$ is invertible.

$\Nc=2$ SYM theory follows this pattern:  one
adds fermionic partners to the fields in the ADHM realization
of  the theory, and the
constraints on the fermionic fields are obtained
by linearizing the constraints eq.~(\ref{c1}) and eq.~(\ref{c2}).
At the Lagrangian level one implements the constraints
via Lagrangian multipliers $H_\C$ and $H_\R$ which multiply eq.~(\ref{c1})
and eq.~(\ref{c2}), and by their fermionic partners $\chi_\R$ and $\chi_\C$
which are then regarded as additional fields.
In \cite{BFMT}, whose notation we follow, the reader will find a detailed analysis
of this case. After regularizing the moduli space of gauge connections by minimally resolving
the singularities, the BRST transformations of the theory lead to
\begin{eqnarray}
\hat\xi^\ast &=& (\phi I-Ia) \vf{}{I} +  (-J\phi+aJ+\epsilon J)\vf{}{J} +
([\phi,B_\ell]+\epsilon_\ell)\vf{}{B_\ell} \nonumber \\
&+& [\phi,H_\R]\vf{}{H_\R} + ([\phi,H_\C]+\epsilon H_\C)\vf{}{H_\R} +
[\phi,\bar\phi]\vf{}{\bar\phi}
\nonumber \\
&+& (\phi\mu_I-\mu_Ia) \vf{}{\mu_I} +  (-\mu_J\phi+a\mu_J+\epsilon\mu_J)\vf{}{\mu_J}
+([\phi,M_\ell]+\epsilon_\ell M_\ell)\vf{}{M_l} \nonumber \\
&+& [\phi,\chi_\R] \vf{}{\chi_\R} +  ([\phi,\chi_\C] +\epsilon\chi_\C) \vf{}{\chi_\R}
+[\phi,\eta]\vf{}{\eta}
\end{eqnarray}
and
\begin{eqnarray}
Q_\xi  &=& \mu_I\vf{}{I}  + \mu_J\vf{}{J} + M_\ell \vf{}{B_\ell}
+[\phi,\chi_\R]\vf{}{H_\R} + \left( [\phi,\chi_\C]+\epsilon
\chi_\C\right)\vf{}{H_\C}+\eta\vf{}{\bar\phi}\nonumber
\\ &+& (\phi I-Ia) \vf{}{\mu_I} + (-J\phi+aJ+\epsilon J)\vf{}{\mu_J} +
([\phi,B_\ell]+\epsilon_\ell B_\ell)\vf{}{M_\ell}\nonumber \\ &+& H_\R\vf{}{\chi_\R}
+ H_\C\vf{}{\chi_\C} +[\phi,\bar\phi]\vf{}{\eta}. \label{BRST2-2}
\end{eqnarray}
(One also includes an ``auxiliary''  bosonic field $\bar\phi$ with its partner $\eta$.)
This new vector field $Q_\xi$ satisfies the conditions of Definition  \ref{defBRST}, so that \emph{for the BRST vector field of
the  ADHM formulation of  $\Nc=2$ SYM, the superlocalization
formula eq.~(\ref{superlocform}) holds.} The reader will find in \cite{BFMT} the
evaluation of the superdeterminant which arises from the application of eq.~(\ref{superlocform}).

. 

\section*{Acknowledgements}
This work was supported in part by the EEC contracts HPRN-CT-2000-00122,
 HPRN-CT-2000-00131  and
HPRN-CT-2000-00148, by the INTAS contract 99-0-590 and  by the MIUR-COFIN
contracts 2001-025492 and 2000-02262971. This paper was mostly
written while the first author was visiting the Tata Institute
for Fundamental Research in Mumbai, Department of Theoretical Physics,
to which thanks are due for hospitality and support.


\begin{thebibliography}{10}\frenchspacing

\bibitem{Amati:1988ft}
D. Amati, Konishi, Y. Meurice, G. C. Rossi, and G. Veneziano.
\newblock Nonperturbative aspects in supersymmetric gauge theories.
\newblock {\em Phys. Rept.}, {\bf 162} (1988) 169--248.
\bibitem{ADHM}
M. F. Atiyah, N. J. Hitchin, V. G. Drinfel'd, and Y. I. Manin.
\newblock Construction of instantons.
\newblock {\em Phys. Lett. A}, {\bf 65} (1978) 185--187.

\bibitem{AHS}
M. F. Atiyah, N. J. Hitchin, and I. M. Singer.
\newblock Self-duality in four-dimensional {R}iemannian geometry.
\newblock {\em Proc. Roy. Soc. London Ser. A}, {\bf 362} (1978) 425--461.

\bibitem{BHR}
C. Bartocci, U. Bruzzo, and D. Hern{\'a}ndez~Ruip{\'e}rez.
\newblock {\em The geometry of supermanifolds}, volume~71 of {\em Mathematics
  and its Applications}.
\newblock Kluwer Academic Publishers Group, Dordrecht, 1991.

\bibitem{Batch}
M. Batchelor.
\newblock The structure of supermanifolds.
\newblock {\em Trans. Amer. Math. Soc.}, {\bf 253} (1979) 329--338.

\bibitem{Bellisai:2000bc}
D. Bellisai, F. Fucito, A. Tanzini, and G. Travaglini.
\newblock Instanton calculus, topological field theories and {N}=2 super
  {Y}ang-{M}ills theories.
\newblock {\em J. High Energy Phys.}, {\bf 07} (2000) 017.

\bibitem{Bellisai:2000tn}
D. Bellisai, F. Fucito, A. Tanzini, and G. Travaglini.
\newblock Multi-instantons, supersymmetry and topological field theories.
\newblock {\em Phys. Lett.}, {\bf B480} (2000) 365--372.

\bibitem{B}
F. A. Berezin.
\newblock {\em Introduction to superanalysis}. \newblock D. Reidel Publishing Co., Dordrecht, 1987.

\bibitem{BL}
F. A. Berezin and D. A. Le\u{\i}tes.
\newblock Supermanifolds.
\newblock {\em Dokl. Akad. Nauk SSSR}, {\bf 224} (1975) 505--508.

\bibitem{BGV}
N. Berline, E. Getzler, and M. Vergne.
\newblock {\em Heat kernels and {D}irac operators}.
\newblock Springer-Verlag, Berlin, 1992.

\bibitem{BFMT}
U. Bruzzo, F. Fucito, J. F. Morales, and A. Tanzini.
\newblock Multi-instanton calculus and equivariant cohomology.
\newblock {\em J. High. Energy. Phys.} {\bf 05} (2003) 054.

\bibitem{BFTT}
U. Bruzzo, F. Fucito, A. Tanzini, and G. Travaglini.
\newblock On the multi-instanton measure for super {Y}ang-{M}ills theories.
\newblock {\em Nucl. Phys.}, {\bf B611} (2001) 205--226.

\bibitem{DK}
S. Donaldson and P. Kronheimer.
\newblock {\em Geometry of Four Manifolds}.
\newblock Oxford Mathematical Monographs. Clarendon Press, Oxford, 1990.

\bibitem{Dorey:2002ik}
N. Dorey, T. J. Hollowood, V. V. Khoze, and M. P. Mattis.
\newblock The calculus of many instantons.
\newblock {\em Phys. Rept.}, {\bf 371} (2002) 231--459.

\bibitem{Dorey:1996hu}
N. Dorey, V. V. Khoze, and M. P. Mattis.
\newblock Multi-instanton calculus in {N}=2 supersymmetric gauge theory.
\newblock {\em Phys. Rev.}, {\bf D54} (1996) 2921--2943.

\bibitem{Dorey:1996bf}
N. Dorey, V. V. Khoze, and M. P. Mattis.
\newblock Multi-instanton calculus in {N}=2 supersymmetric gauge theory.
  {I}{I}: Coupling to matter.
\newblock {\em Phys. Rev.}, {\bf D54} (1996) 832--848.

\bibitem{Flume:2002az}
R. Flume and R. Poghossian.
\newblock An algorithm for the microscopic evaluation of the coefficients of
  the {S}eiberg-{W}itten prepotential,
\newblock {\tt hep-th/0208176.}

\bibitem{Fucito:2001ha}
F. Fucito, J. F. Morales, and A. Tanzini.
\newblock D-instanton probes of non-conformal geometries.
\newblock {\em J. High Energy Phys.}, {\bf 07} (2001) 012--040.

\bibitem{Fucito:1997ua}
F. Fucito and G. Travaglini.
\newblock Instanton calculus and nonperturbative relations in {N}=2
  supersymmetric gauge theories.
\newblock {\em Phys. Rev.}, {\bf D55} (1997) 1099--1104.

\bibitem{HM}
D. Hern{\'a}ndez~Ruip{\'e}rez and J. Mu{\~n}oz~Masqu{\'e}.
\newblock Construction intrins\`eque du faisceau de {B}erezin d'une vari\'et\'e
  gradu\'ee.
\newblock {\em C. R. Acad. Sci. Paris S\'er. I Math.}, {\bf 301} (1985) 915--918.

\bibitem{Hollowood:2002ds}
T. J. Hollowood.
\newblock Calculating the prepotential by localization on the moduli space of
  instantons.
\newblock   {\em J. High Energy Phys.}, {\bf 03} (2002) 038--061.

\bibitem{Hollowood:2002zv}
T. J. Hollowood.
\newblock Testing {S}eiberg-{W}itten theory to all orders in the instanton
  expansion.
\newblock {\em Nucl. Phys.}, {\bf B639} (2002) 66--94.

\bibitem{J}
A. Johansen.
\newblock `Twisting of $N=1$ SUSY gauge theories and heterotic topological theories.
\newblock {\em Inernat. J. Mod.  Phys. A} {\bf 10} (1995) 4325-4357.


\bibitem{K}
B. Kostant.
\newblock Graded manifolds, graded {L}ie theory, and prequantization.
\newblock In {\em Differential geometrical methods in mathematical physics
  (Proc. Sympos., Univ. Bonn, Bonn, 1975)}, pp. 177--306. Lecture Notes in
  Math. {\bf 570}. Springer, Berlin, 1977.

\bibitem{Lerche:1997sm}
W. Lerche.
\newblock Introduction to {S}eiberg-{W}itten theory and its stringy origin.
\newblock In {\em Trends in theoretical physics (La Plata, 1997)},
   {\em AIP Conf. Proc.} {\bf 419}, pp. 171--217. Amer. Inst. Phys., Woodbury, NY,
  1998.

\bibitem{Losev:1998tp}
A. Losev, N. Nekrasov, and S. L. Shatashvili.
\newblock Issues in topological gauge theory.
\newblock {\em Nucl. Phys. B}, {\bf 534} (1998) 549--611.



\bibitem{M}
N. Marcus.
\newblock The other topological twisting of N=4 Yang-Mills.
\newblock {\em Nucl.  Phys. B}, {\bf 452} (1995) 331-345.


\bibitem{Moore:1998et}
G. W. Moore, N. Nekrasov, and S. Shatashvili.
\newblock D-particle bound states and generalized instantons.
\newblock {\em Commun. Math. Phys.}, {\bf 209} (2000) 77--95.

\bibitem{Moore:1997dj}
G. W. Moore, N. Nekrasov, and S. Shatashvili.
\newblock Integrating over {H}iggs branches.
\newblock {\em Commun. Math. Phys.}, {\bf 209} (2000) 97--121.

\bibitem{Naka}
H. Nakajima.
\newblock   Lectures on {H}ilbert schemes of points on surfaces,
  {\em University Lecture Series} {\bf 18}.
\newblock American Mathematical Society, Providence, RI, 1999.

\bibitem{Nekrasov:2002qd}
N. A. Nekrasov.
\newblock Seiberg-{W}itten prepotential from instanton counting.
\newblock In {\em Proceedings of the International Congress of Mathematicians,
  Vol. III (Beijing, 2002)}, pp. 477--495, Beijing, 2002. Higher Ed. Press.

\bibitem{Seiberg:1994rs}
N. Seiberg and E. Witten.
\newblock Electric-magnetic duality, monopole condensation, and confinement in
  {N}=2 supersymmetric {Y}ang-{M}ills theory.
\newblock  {\em Nucl. Phys. B}, {\bf B426} (1994) 19--52.

\bibitem{Seiberg:1994aj}
N. Seiberg and E. Witten.
\newblock Monopoles, duality and chiral symmetry breaking in {N}=2
  supersymmetric {QCD}.
\newblock {\em Nucl. Phys. B}, {\bf 431} (1994) 484--550.

\bibitem{Shifman:1994ee}
M. Shifman, ed.
\newblock {\em Instantons in gauge theories}.
\newblock World Scientific Publishing Co. Inc., River Edge, NJ, 1994.

\bibitem{VW}
C Vafa and E. Witten.
\newblock  A strong coupling test of S duality.
\newblock  {\em  Nucl. Phys. B}, {\bf 431} (1994) 3-77.

\bibitem{Witten:1988ze}
E. Witten.
\newblock Topological quantum field theory.
\newblock   {\em Commun. Math. Phys.}, {\bf  117} (1988) 353--386.

\bibitem{Witten95}
E. Witten.
\newblock Supersymmetric {Y}ang-{M}ills theory on a four-manifold.
\newblock  {\em J. Math. Phys.}, {\bf 35} (1994)  5101--5135.

\bibitem{Y}
J. P. Yamron.
\newblock Topological actions from twisted supersymmetric theories.
\newblock {\em Phys. Lett. B}, {\bf 213} (1988) 325-330.

\bibitem{ZJ}
J. Zinn-Justin.
\newblock  Quantum field theory and critical phenomena,  {\em
  International Series of Monographs on Physics} {\bf 85}.
\newblock  Oxford University Press, New York, 1993.

\end{thebibliography}
\end{document}